\documentclass[12pt]{article}

\usepackage[numbers]{natbib}
\usepackage{times}

\begin{document}

\newcommand{\comment}[1]{}

\begin{center}


\bigskip\smallskip

{\Large \bf  \boldmath
$Z$ polarization at an $e^+e^-$ collider and properties of decay-lepton
angular asymmetries
}

\bigskip\smallskip

Kumar Rao$^a$, Saurabh D. Rindani$^b$\footnote{Corresponding author.
Email: sdrindani@gmail.com}, Priyanka Sarmah$^c$, Balbeer
Singh$^d$

\bigskip\smallskip
{\it $^a$Department of Physics, Indian Institute of Technology Bombay,\\
Powai,
Mumbai 400076, India}

\smallskip

{\it $^b$Theoretical Physics Division, Physical Research Laboratory,\\
Navrangpura, Ahmedabad 380009, India
}

\smallskip

{\it $^c$ 
Department of Physics, National Tsing Hua University, 
\\Hsinchu 30013, Taiwan
}
 
\smallskip

{\it $^d$
Department of Physics, University of South Dakota, 
\\Vermillion, SD 57069, USA 
}

\bigskip \smallskip

\end{center}


\newpage

\begin{center}
\noindent{\large \bf Abstract}
\end{center}

$Z$ production at an $e^+e^-$ collider, associated with production of
other particles, can be an accurate source of information of details of
electroweak interactions, including possible interactions beyond the
standard model. This is because of the inherent low background at an 
$e^+e^-$ collider as well as the fact that the $Z$ state can be fully
reconstructed. If the $Z$ decays into a lepton pair, such a final state
in spite of the lower branching ratio, can be studied with great
accuracy. One such process is $e^+e^- \to HZ$, where is $H$ is the Higgs
boson, the most important process for Higgs production at an $e^+e^-$ 
collider.  
Apart from the angular distribution, the polarization of the $Z$, 
fully characterized by its spin density matrix
in the production process, can give detailed information about the production
process. The vector and tensor polarization parameters of the $Z$ density
matrix can be related to angular asymmetries of the decay leptons in the $Z$
rest frame. Thus, an experimental study of these lepton asymmetries can give
information on the underlying interactions in $Z$ production.  
We discuss from a general physical point of view the properties of the 
density matrix as well as lepton angular asymmetries. While most
considerations will be applicable to processes of $Z$ production associated
with any other particle or particles, for some discussions, we specialize to a
$HZ$ final state. 
While many of the results can be found in earlier literature, especially
for the process $e^+e^- \to HZ$, we
give details of the reasoning, which are not always found.
We discuss the properties of the spin density matrix under C, P and T
transformations, and combinations thereof, 
as also the predictions of these for the corresponding 
leptonic asymmetries.
The specific transformations P, CP, T and CPT are of special importance
and we discuss the consequences of these symmetries or their absence for
the leptonic asymmetries. 
A specific issue which has been given attention to is the role of beam
polarization, and how one can infer on general grounds which 
asymmetries get enhanced by the use of beam polarization. Similarly, we
also discuss which asymmetries would be sensitive to the measurement of tau polarization in $Z$ decay in to $\tau^+\tau^-$.

\newpage

\section{Introduction}

The discovery of the Higgs boson ($H$) has put on firm ground the
standard model (SM) of particle physics. 
Nevertheless, the determination of various SM
parameters does not have sufficient precision to nail down the SM as the
completely correct model. 

An important work on relating angular distributions of decay products of
particles of various spins to the production density matrices was done
by Boudjema and Singh \cite{Boudjema:2009fz}. In the specific context of $W$ and $Z$
bosons, the relation between the production vector and tensor
polarizations and the angular distributions and angular asymmetries of
leptons produced in the decays of these bosons was established in a
convenient form by Rahaman and Singh \cite{rahaman}.
This formalism was
applied to various processes in \cite{rahaman,
Rahaman:2017qql,
Rao:2018abz,
Rahaman:2018ujg,
Rao:2019hsp,
Rahaman:2019mnz,
Rahaman:2019lab,
Rahaman:2020jll,
Rao:2020hel}.
An approach relating the
angular distributions to spin-one polarization parameters has only been
pursued recently, starting with the work in \cite{rahaman} (see also
\cite{Nakamura:2017ihk, Aguilar-Saavedra:2017zkn}).
The determination of $HZZ$ and $HWW$ couplings, as also
of triple-gauge couplings through decay-lepton angular distributions 
 has been addressed in the past in several works 
(\cite{hagistong, Hagiwara:2000tk, Biswal:2005fh, Godbole:2007cn, 
Biswal:2008tg, Biswal:2009ar, Rindani:2009pb, Rindani:2010pi,
Godbole:2014cfa, Godbole:2013lna, Craig:2015wwr, 
Beneke:2014sba,  Khanpour:2017cfq, Zagoskin:2018wdo, Li:2019evl, 
He:2019kgh} and references therein),
 
Experimental determination of the Higgs couplings has been carried out 
at Large Hadron Collider (LHC) experiments. They have in general been 
found
consistent with the predictions of the SM. The search
is on for possible deviations from SM couplings, or the so-called 
anomalous couplings. 

The ATLAS and CMS collaborations working at the LHC 
have recently studied Higgs boson couplings to
$WW$ and $ZZ$ and presented results using the vector-boson fusion
mechanism for Higgs production, as well as using Higgs decay 
into a vector-boson pair \cite{ATLAS:2021pkb,
CMS:2019ekd,
CMS:2019jdw,
ATLAS:2020evk,
CMS:2018zzl,
ATLAS:2018jvf,
ATLAS:2018xbv}.

While future LHC experiments will succeed in reducing the
uncertainties in the determination of the anomalous couplings, an
electron-positron collider at centre-of-mass (cm) energy of a few 
hundred GeV to a few TeV, proposed to be  constructed, would fare 
much better than
the LHC. Several of the papers cited above are concerned with
$e^+e^-$ colliders, and they have discussed the expected sensitivities
of various experiments and their configurations. While we will review
them in passing, the purpose of this work is to elaborate on the 
physical
understanding of the proposals to study anomalous couplings through
angular distributions in $Z$ decay. 

Using the relations derived between the elements of the $Z$ spin density
matrix, parameterized by the vector and tensor polarization parameters,
and the angular distribution of charged leptons in a specific frame,
following \cite{rahaman}, we examine several general expectations
following from C, P and T transformation properties of the density
matrix elements. Thus, we will try to predict on general grounds 
which asymmetries would be
sensitive to CP violating couplings and which would not be, which
asymmetries would isolate absorptive part of the production amplitude,
which asymmetries would be potentially enhanced by longitudinal
polarization of electron and positron beams, and which asymmetries
would be enhanced by the measurement of the fermion polarization in the
final state.

While most of the physical ideas will apply
to more general processes with inclusive production of a spin-one
particle, we restrict ourselves to a class of 
processes, with a $Z$ produced in electron-positron collisions. 
One reason for restricting to $Z$ and not considering $W$ production is
the possibility of reconstructing the $Z$ momentum from the charged
lepton momenta produced in its decay, whereas $W$ momentum would be
difficult to reconstruct. 
The general process we study is $e^+e^- \to Z +X $, where $X$ 
is typically a one-particle or a two-particle state. 
Further, the processes most easily
categorized for studying beam polarization effects are those with an
$s$-channel exchange of a single boson ($\gamma$ or $Z$). The most
important of such processes is $e^+e^- \to HZ$, the dominant process for
Higgs production at an electron-positron collider. 
Specifically, we will concentrate on the
charged lepton asymmetries in $Z$ decay in the process 
$e^+e^- \to HZ$, to illustrate our ideas.

The plan of the paper is as follows. In the next section we describe the
formalism proposed in \cite{rahaman} for calculating angular asymmetries of
charged leptons in $Z$ decay by first writing down the spin density 
matrix
for the $Z$ in terms of the helicity amplitudes for the
production process, and relating the density matrix elements to the
polarization parameters, which are simply related to angular 
asymmetries. 
We next discuss in Sec. 3 how the asymmetries can be classified in terms of the 
C, P and T properties, and how general inferences regarding the 
underlying
interactions can be drawn from these properties. In the following 
section we talk about CPT-odd asymmetries and their relation to
absorptive parts of amplitudes. In Sec. 5 we
apply the general discussions of the earlier sections to the 
concrete case of the process $e^+e^- \to HZ$, giving explicit
expressions for the density matrix and asymmetries. We also address
 the question of the
role of beam polarization, and which asymmetries are enhanced by
polarization, specifically in the $HZ$ production process.
The final section (Sec. 6) contains conclusions and a discussion of the
results.

\section{\boldmath The $Z$ spin density matrix and decay lepton 
asymmetries}

We first need to calculate the spin density matrix for the $Z$, for which we
will need to evaluate the helicity amplitudes for the process. We can then
write the density matrix for the process $e^+e^- \to Z + X$ as 
 \begin{equation}\label{denmat}
\displaystyle \rho(i,j)= \frac{1}{4}
\sum_{\lambda,\lambda^{'},h}M(\lambda,\lambda^{'},i,h)M^{\ast}(\lambda,\lambda^{'},j,h).
 \end{equation}
Here $i,j$ refer to the helicities $-,0,+$ of the $Z$, whereas 
$\lambda,\lambda'$ to
the (signs of the) helicities $-,+$ of the electron and positron,
respectively. $h$ refers collectively to the helicities of the 
particles in the state $X$. 

This assumes unpolarized electron and positron beams. However, we would
also be interested in polarized beams. Denoting by $P_L$ and $\bar P_L$
respectively the degrees of longitudinal polarization of the electron
and positron, the density matrix including beam polarization is given by
 \begin{equation}\label{denmatpol}
\displaystyle \rho(i,j)=\frac{1}{4}
\sum_{\lambda,\lambda^{'},h} (1+\lambda P_L)(1+\lambda' \bar P_L)
M(\lambda,\lambda^{'},i,h )M^{\ast}(\lambda,\lambda^{'},j,h )
 \end{equation}

Following \cite{rahaman}, we  now show how decay-lepton angular
asymmetries may be written in terms of the $Z$ spin density matrix
defined above. 

Defining an integral of this density matrix over an appropriate kinematic
range as $\sigma(i,j)$, the latter can be parameterized in terms of the linear 
polarization $\vec P$ and the tensor polarization $T$ as follows 
\cite{Leader:2001gr}.
\begin{equation}\label{vectensorpol}
\sigma(i,j) \equiv \sigma \left( 
\begin{array}{ccc}
\frac{1}{3} + \frac{P_z}{2} + \frac{T_{zz}}{\sqrt{6}} &
\frac{P_x - i P_y}{2\sqrt{2}} + \frac{T_{xz}-i T_{yz}}{\sqrt{3}} &
\frac{T_{xx}-T_{yy}-2iT_{xy}}{\sqrt{6}} \\
\frac{P_x + i P_y}{2\sqrt{2}} + \frac{T_{xz}+i T_{yz}}{\sqrt{3}} &
\frac{1}{3} - \frac{2 T_{zz}}{\sqrt{6}} &
\frac{P_x - i P_y}{2\sqrt{2}} - \frac{T_{xz}-i T_{yz}}{\sqrt{3}} \\
\frac{T_{xx}-T_{yy}+2iT_{xy}}{\sqrt{6}} &
\frac{P_x + i P_y}{2\sqrt{2}} - \frac{T_{xz}+i T_{yz}}{\sqrt{3}} &
\frac{1}{3} - \frac{P_z}{2} + \frac{T_{zz}}{\sqrt{6}} 
\end{array}
\right)
\end{equation}
where $\sigma(i,j)$ is the integral of $\rho(i,j)$, and $\sigma$ is the 
production cross section, 
\begin{equation}\label{totalcs}
\sigma = \sigma(+,+) + \sigma(-,-) + \sigma(0,0).
\end{equation}
We now construct the polarization parameters from
the integrated density matrix elements and give relations of these to 
angular asymmetries of the decay leptons which would serve as measures
of the polarization parameters. 

The eight independent vector and tensor polarization observables can be
extracted using appropriate linear combinations of the integrated
density matrix elements of eq. (\ref{vectensorpol}):
 \begin{eqnarray}\label{pol1}
P_{x}&=&  \frac{\lbrace \sigma(+,0)+\sigma(0,+)
 + \sigma(0,-)+\sigma(-,0)\rbrace}{\sqrt{2}\sigma}\\\label{pol2}
 P_{y}&=&\frac{- i  \lbrace[\sigma(0,+)-\sigma(+,0)]+[\sigma(-,0)-\sigma(0,-)]\rbrace}{\sqrt{2}\sigma}\\\label{pol3}
 P_{z}&=&\frac{[\sigma(+,+)]-[\sigma(-,-)]}{\sigma}\\\label{pol4}
 T_{xy}&=&\frac{- i  \sqrt{6}[\sigma(-,+)-\sigma(+,-)]}{4\sigma}\\\label{pol5}
  T_{xz}&=&\frac{\sqrt{3}\lbrace[\sigma(+,0)+\sigma(0,+)]-[\sigma(0,-)+\sigma(-,0)]\rbrace}{4\sigma}\\\label{pol6}
  T_{yz}&=&\frac{- i  \sqrt{3}\lbrace[\sigma(0,+)-\sigma(+,0)]-[\sigma(-,0)-\sigma(0,-)]\rbrace}{4\sigma}\\\label{pol7}
  T_{xx}-T_{yy}&=&\frac{\sqrt{6}[\sigma(-,+)+\sigma(+,-)]}{2\sigma}\\\label{pol8}
  T_{zz}&=&
\frac{\sqrt{6}[\sigma(+,+)+\sigma(-,-)-2 \sigma(0,0)]}{6\sigma}
\label{pol9}
 \end{eqnarray}
Of these $P_x$, $P_y$ and $P_z$ are the vector polarizations, whereas
the $T$'s are the tensor polarizations, with the constraint that the
tensor is traceless.

Experimentally, the spin information of the $Z$ is obtained from kinematic
distributions of its decay products. 
Ref. \cite{rahaman} describes
  the formalism that connects 
all the spin observables of $Z$  to the angular distribution of the leptons 
arising from its decay, and further defines
 various asymmetries 
in the rest frame of the $Z$ which are simply related to
the polarization observables given in eqs. (\ref{pol1})-(\ref{pol9}). 
These are given by
\begin{equation}\label{asyx}
A_{x}\equiv\frac{\sigma(\cos\phi^{\ast}>0)-\sigma(\cos\phi^{\ast}<0)}{\sigma(\cos\phi^{\ast}>0)+\sigma(\cos\phi^{\ast}<0)}=\frac{3\alpha P_{x}}{4}
\end{equation}
\begin{equation}\label{asyy}
A_{y}\equiv\frac{\sigma(\sin\phi^{\ast}>0)-\sigma(\sin\phi^{\ast}<0)}{\sigma(\sin\phi^{\ast}>0)+\sigma(\sin\phi^{\ast}<0)}=\frac{3\alpha P_{y}}{4}
\end{equation}
\begin{equation}\label{asyz}
A_{z}\equiv\frac{\sigma(\cos\theta^{\ast}>0)-\sigma(\cos\theta^{\ast}<0)}{\sigma(\cos\theta^{\ast}>0)+\sigma(\cos\theta^{\ast}<0)}=\frac{3\alpha P_{z}}{4}
\end{equation}
\begin{equation}\label{asyxy}
A_{xy}\equiv\frac{\sigma(\sin2\phi^{\ast}>0)-\sigma(\sin2\phi^{\ast}<0)}{\sigma(\sin2\phi^{\ast}>0)+\sigma(\sin2\phi^{\ast}<0)}
=\frac{2}{\pi}\sqrt{\frac{2}{3}}T_{xy}
\end{equation}
\begin{equation}\label{asyxz}
A_{xz}\equiv\frac{\sigma(\cos\theta^{\ast}\cos\phi^{\ast}<0)-\sigma(\cos\theta^{\ast}\cos\phi^{\ast}>0)}{\sigma(\cos\theta^{\ast}\cos\phi^{\ast}>0)+\sigma(\cos\theta^{\ast}\cos\phi^{\ast}<0)}
=\frac{-2}{\pi}\sqrt{\frac{2}{3}}T_{xz}
\end{equation}
\begin{equation}\label{asyyz}
A_{yz}
\equiv\frac{\sigma(\cos\theta^{\ast}\sin\phi^{\ast}>0)-\sigma(\cos\theta^{\ast}\sin\phi^{\ast}<0)}{\sigma(\cos\theta^{\ast}\sin\phi^{\ast}>0)+\sigma(\cos\theta^{\ast}\sin\phi^{\ast}<0)}
=\frac{2}{\pi}\sqrt{\frac{2}{3}}T_{yz}
\end{equation}
\begin{equation}\label{asyx2y2}
A_{x^{2}-y^{2}}
\equiv\frac{\sigma(\cos2\phi^{\ast}>0)-\sigma(\cos2\phi^{\ast}<0)}{\sigma(\cos2\phi^{\ast}>0)+\sigma(\cos2\phi^{\ast}<0)}
=\frac{1}{\pi}\sqrt{\frac{2}{3}}(T_{xx}-T_{yy})
\end{equation}
\begin{equation}\label{asyzz}
A_{zz}
\equiv\frac{\sigma(\sin3\theta^{\ast}>0)-\sigma(\sin3\theta^{\ast}<0)}{\sigma((\sin3\theta^{\ast}>0)+\sigma((\sin3\theta^{\ast}<0)}
=\frac{3}{8}\sqrt{\frac{3}{2}}T_{zz}
\end{equation}
Here, $\alpha$ is the $Z$ boson polarization analyser, given in terms 
of its left- and right-handed couplings to charged leptons, respectively $L_\ell$ and $R_\ell$, by
\begin{equation}\label{polanalyser}
\alpha =\frac{R_\ell^2 -L_\ell ^2}{R_\ell^2 +L_\ell ^2}.
\end{equation}
$\alpha$ may also be written in terms of the vector and axial-vector
couplings $c_V$ and $c_A$ as
\begin{equation}\label{alpha}
\alpha = - \frac{2 c_V c_A}{c_V^2 +c_A^2}
\end{equation}
The angles $\theta^\ast$ and $\phi^\ast$ are polar and azimuthal angles of
the lepton in the rest frame of the $Z$. This frame is reached by a
combination of a boost and a rotation from the laboratory frame. In the laboratory frame, the initial
$e^-$ beam defines the $z$ axis, and the production plane of $Z$ is
defined as the $xz$ plane. 
While boosting to the $Z$ rest frame, the $xz$
plane is kept unchanged. Then, the angles $\theta^\ast$ and $\phi^\ast$
are  measured with respect to the
would-be momentum of the $Z$. 

Details of the transformations involved are as follows.
We start with the lab. frame $F$, in which  $$\hat p_{e^-} = - \hat
p_{e^+} = \hat z,$$ and the components of the $Z$ momentum 
three-vector, lying in the $xz$ plane, are
\begin{equation}
\vec k_Z  =\vert\vec k_Z\vert (\sin\theta, 0, \cos\theta).
\end{equation}

The next  frame we consider is $F'$ obtained from $F$ by rotation about
the $y$ axis, so that $\vec k_Z$ lies along the new $z'$ axis, 
(with $y'\equiv y$). The components of the $Z$ three-momentum are given by
\begin{equation}
\vec k_Z' = \vert \vec k_Z \vert (0,0,1). 
\end{equation}
In $F'$, the three-momentum
 of the incoming electron is given by
\begin{equation}
\vec p\,_{e^-}' = \frac{\sqrt{s}}{2} (-\sin\theta, 0, \cos\theta ).
\end{equation}

We next boost the $Z$ momentum so that it comes to rest. In this frame $F''$,
$\vec k_Z'' = (0,0,0)$ and $ E_Z'' = m_Z$.
Also, the momenta of the charged leptons from $Z$ decay are
\begin{equation}
\vec p\,_{\ell^-}'' = - \vec p\,_{\ell^+}'' = E''_{\ell^-} (\sin\theta''_{\ell^-}
\cos\phi''_{\ell^-},
\sin\theta''_{\ell^-} \sin\phi''_{\ell^-}, \cos\theta''_{\ell^-}).
\end{equation}
 The $x''z''$ 
plane 
coincides with $x'z'$ as also the $xz$ plane. That is, 
the $y$ and $y''$ directions coincide. So 
\begin{equation}
 E''_{\ell^-}\sin\theta''_{\ell^-} \sin\phi''_{\ell^-} = E'_{\ell^-}
\sin\theta'_{\ell^-}\sin\phi'_{\ell^-}
\end{equation}

After boost by the $Z$ velocity, $\vec p\,_{\ell^-}''$ is given by 
\begin{equation}
\vec p\,_{\ell^-}'' = E'_{\ell^-} \left(\sin\theta'_{\ell^-} \cos\phi'_{\ell^-},
\sin\theta'_{\ell^-} \sin\phi'_{\ell^-},
\frac{(\cos\theta'_{\ell^-} - \beta_Z)}{\sqrt{1-\beta_Z^2}} \right)
\end{equation}
\begin{equation}E''_{\ell^-} = E'_{\ell^-}\frac{(1  - \beta_Z
\cos\theta_{\ell^-}')}{\sqrt{1-\beta_Z^2}}
\end{equation}
Note that the angles $\theta''_{\ell^-}$ and $\phi''_{\ell^-}$ are the same as
the angles $\theta^\ast$ and $\phi^\ast$ used earlier in defining the
asymmetries.

\section{C, P, T properties of the lepton 
asymmetries}

Here we study transformation properties under discrete transformations
C, P, T and their combinations of the angular asymmetries,
which are defined in terms of momentum three-vectors of various
particles entering the process. We will therefore not be concerned with
transformation properties of the amplitudes as such, and the intrinsic
phases appearing in the C, P and T transformations of the fields will
not play a role.

The discussion here is valid for all processes $e^+e^- \to Z +X$, and not
restricted to the Higgs production process $e^+e^- \to HZ$.
We shall use T to mean naive time reversal, that is,
reversal of the signs of all momenta and spins. We will not be concerned
with genuine time-reversal operation which also implies interchange of
initial and final states, which is difficult to achieve in practice.

Let us now examine the properties of various momenta and axes under the
transformations of P, T and C, with a view to determining the
transformation properties of the angles of the decay leptons, and hence
the transformations of the asymmetries.

Under P, all momenta change sign. Since the $xz$ plane is defined by the
$e^-$ and $Z$ momenta, both of which flip sign under P, the $y$ axis,
which may be thought of as $\hat z \times \hat x$, is unchanged. Since
the $y$ axis is the same in $F'$ as also in $F''$, it is unchanged under
P even in $F''$. However, in $F''$, 
$\vec p\,_{\ell^-}$ changes sign, hence its $y$ component changes sign. 
The $x$ and $z$ axes reverse directions,
but since $\vec p\,_{\ell^-}$ also changes sign, its $x$ and $z$ 
components are unchanged under P. Thus, under P, 
 $\cos\theta''_{\ell^-} \to
\cos\theta''_{\ell^-}$, $\cos\phi''_{\ell^-} \to \cos\phi''_{\ell^-}$,
$\sin\phi''_{\ell^-} \to - \sin\phi''_{\ell^-}$. In other words, 
$\theta''_{\ell^-} \to
\theta''_{\ell^-}$, $\phi''_{\ell^-} \to 2\pi - \phi''_{\ell^-}$.

The transformation properties under T are the same as those under P.

Thus, under P and T, asymmetries which are labeled with a single $y$
suffix change sign, whereas the others are invariant. This means that
$A_y$, $A_{xy}$ and $A_{yz}$ change sign.

As for C, the $e^-$ and $e^+$ momenta are interchanged, which in the
frame $F$ amounts a change of sign for each of these, since they are
equal in magnitude and oppositely directed. Thus the $z$ axis in $F$ 
changes sign under C. The momentum of the $Z$ does not change under C.
For the sake of definiteness, let us define the $x$ axis such that the
$Z$ momentum lies in the $xz$ plane, and the $x$ component of the $Z$
momentum is positive. Under C, the $z$ axis changes sign, whereas the
$x$ axis does not. Hence the $y$ axis defined as $\hat z \times \hat x$
changes direction. 

Again, in the frame $F''$, the momenta of $\ell^-$ and $\ell^+$ are
interchanged under C, and since they are equal and opposite, they change
sign under C. However, since the $z$ axis is unchanged under C, being
along the momentum of the $Z$ in $F'$, the $z$ components of $\ell^-$
and $\ell^+$ change sign, resulting in $\cos\theta''_{\ell^-} \to -
\cos\theta''_{\ell^-}$, and $\theta''_{\ell^-} \to
\pi - \theta''_{\ell^-}$.  Since the $y$ axis as also
the $\ell^{-}$ momentum change direction, the $y$ component of the
$\ell^{-}$ remains unchanged. Similarly, the $x$ axis direction, given
by $\hat y \times \hat z$, changes direction since $\hat y$ changes
sign and $\hat z$ does not. Since $\ell^-$ flips direction, as well as
the $x$ axis, the $x$ component does not change sign. 
Thus, under C, $\cos\phi''_{\ell^-} \to \cos\phi''_{\ell^-}$;
$\sin\phi''_{\ell^-} \to \sin\phi''_{\ell^-}$ and, equivalently, 
$\phi''_{\ell^-} \to \phi''_{\ell^-}$.  

It follows that under C, those asymmetries with a single $z$ suffix
change sign, the others remaining unchanged. Thus $A_{z}$, $A_{xz}$ and
$A_{yz}$ change sign under C.

From the above, it is clear that under a combination CP of C and P,
$\theta''_{\ell^-} \to \pi - \theta''_{\ell^-}$ and $\phi''_{\ell^-} \to
2\pi - \phi''_{\ell^-}$. 
In other words, $z$, $y$ components of $\vec p\,_{\ell^-}$ change sign 
under CP, but not the $x$ component.
Hence the asymmetries which flip sign under CP are $A_{xy}$ and
$A_{xz}$.

Finally, combining the information gathered above, under CPT, 
the $z$ component of $\vec p\,_{\ell^-}$ changes sign, not the $x$ and $y$
components. The asymmetries which are odd under CPT are thus the same as
 those which are odd under C, {\it viz.}, $A_{z}$, $A_{xz}$ and
$A_{yz}$.

Table \ref{AsyCPT} lists the various transformation properties of the
asymmetries listed
above. The second line of the heading in the table lists the asymmetries in the
notation of Hagiwara and Stong \cite{hagistong}. 

\begin{table}
\centering
\begin{tabular}{|l|c|c|c|c|c|c|c|c|}
\hline
&  $A_{x}$ & $A_{y}$ & $A_{z}$ & $A_{xy}$ & $A_{xz}$ & $A_{yz}$& $A_{zz}$&
$A_{x^2-y^2}$  \\
\hline
&    $A_4$ & $A_7$ & $A_3$ & $A_9$ & $A_5$ & $A_8$ & $A_{12}$ &$A_6$ \\
\hline
P&$    +1$&$    -1$&$    +1$&$    -1$&$  +1$&$   -1$&$    +1$&$
+1$\\
T&$    +1$&$    -1$&$    +1$&$    -1$&$  +1$&$   -1$&$    +1$&$
+1$\\
C&$    +1$&$    +1$&$    -1$&$    +1$&$  -1$&$   -1$&$    +1$&$
+1$\\
CP&$   +1$&$    -1$&$    -1$&$    -1$&$  -1$&$   +1$&$    +1$&$
+1$\\
CPT&$  +1$&$    +1$&$    -1$&$    +1$&$  -1$&$   -1$&$    +1$&$
+1$\\
\hline 
\end{tabular}
\caption{The transformation properties 
of the asymmetries $A_i$ 
under C, P, T, 
 and their combinations.
}\label{AsyCPT}
\end{table}

\subsection{Additional asymmetries}

We can construct a new set of asymmetries $A'_i$ 
by combining each of the above
 asymmetries $A_i$ with the $Z$
forward-backward (FB) asymmetry.
These asymmetries have been used in earlier work \cite{
Rahaman:2019mnz, Rahaman:2019lab, hagistong, Nakamura:2018bli}.

The new asymmetries $A'_i$ can be defined just as $A_i$ are defined in
eqs. (\ref{asyx}) - (\ref{asyzz}), except that each term in 
the numerator in each
of these equations will be a difference of integrals over the c.m. $Z$ 
polar angle $\theta$ from $0$ to $\pi/2$ and from $\pi/2$ to $\pi$. The
denominator will continue to be an integral over $\theta$ over the full
range $0$ to $\pi$. For
example, denoting by $\sigma_{\rm F}$ and $\sigma_{\rm B}$ the cross sections with
$\cos\theta>0$ and $\cos\theta<0$, respectively, 
 $A'_x$ can be written as 
\begin{equation}\label{asyxprime}
A'_{x}\equiv\frac{\sigma_{\rm F}(\cos\phi^{\ast}>0)
-\sigma_{\rm B}(\cos\phi^{\ast}>0)
-\sigma_{\rm F}(\cos\phi^{\ast}<0)
+\sigma_{\rm B}(\cos\phi^{\ast}<0)}
{\sigma(\cos\phi^{\ast}>0)+\sigma(\cos\phi^{\ast}<0)}.
\end{equation}
The other $A'_i$ can be written similarly.

In this case, the C, P, T properties of the new asymmetries change as
compared to the properties of the original symmetries 
because the $Z$ FB asymmetry characterized by the sign of 
$\vec p_Z \cdot (\vec p_{e^-} - \vec p_{e^+})$
is C odd, P even, T even, CP odd and CPT odd, as can be seen from this
equation. 

It may be worth reminding the reader that unlike a (active)
parity transformation usually understood as a transformation
$\theta \rightarrow \pi - \theta$ of the polar coordinate $\theta$,
when the $z$ axis is fixed in space, the parity transformation
here reverses not only $\vec p_Z$ but also  $(\vec p_{e^-} -
\vec p_{e^+})$, whose direction defines the $z$ axis, itself changes
direction. As a result, the FB asymmetry is even under P (as well as T).

The C, P and T properties of $A'_i$ are given in 
Table \ref{A'}.
The second line of the heading in the table lists the asymmetries in the
notation of \cite{hagistong}.
(Note that according to Hagiwara and Stong \cite{hagistong}, $A_9'$ is
zero, which we will see to be true for our Higgsstrahlung process, and that 
they do not list $A_3'$).

\begin{table}[htb]
\centering
\begin{tabular}{|l|c|c|c|c|c|c|c|c|}
\hline
   &  $A_{x}'$&   $A_{y}'$&  $A_{z}'$ & $A_{xy}'$& $A_{xz}'$&  $A_{yz}'$&
$A_{zz}'$&
$ A_{x^2-y^2}'$ \\
\hline
    & $A_4'$ &  $A_7'$ &  $A_3'$ &  $A_9'$ & $A_5'$ & $A_8'$ & $A_{12}'$&$A_6'$ \\
\hline
P     &$+1$    &$-1$    &$+1$    &$-1$  &$+1$   &$-1$   &$+1$    &$+1$\\
T     &$+1$    &$-1$    &$+1$    &$-1$  &$+1$   &$-1$   &$+1$    &$+1$\\
C     &$-1$    &$-1$    &$+1$    &$-1$  &$+1$   &$+1$   &$-1$    &$-1$\\
CP    &$-1$    &$+1$    &$+1$    &$+1$  &$+1$   &$-1$   &$-1$    &$-1$\\
CPT   &$-1$    &$-1$    &$+1$    &$-1$  &$+1$   &$+1$   &$-1$    &$-1$\\
\hline 
\end{tabular}
\caption{The transformation properties under C, P, T, 
 and their combinations,
of the asymmetries $A'_i$ formed by combining the asymmetries $A_i$ listed in
Table \ref{AsyCPT} with the $Z$ forward-backward asymmetry.}\label{A'}
\end{table}

\section{CPT-odd asymmetries and the presence of absorptive amplitudes}

As mentioned earlier, by T we mean not genuine time reversal, but naive
time reversal, under which all momenta and spins change sign.

Genuine time reversal takes a process $i\to f$ 
to a process $f^T \to  i^T$ (where T means that spins and momenta are 
reversed in sign). This is why T operator is anti-unitary. However, 
the asymmetries we are talking about are for the process $i \to  f$ 
with spin and momenta reversed, not $f \to  i$. But if we do not have 
loops/absorptive parts, the forward and backward processes have the 
same amplitude apart from a phase because in the unitarity relation:
\begin{equation}
\langle f |M |i \rangle - \langle i|M^*| f \rangle = i \sum_m 
\langle f |M | m \rangle \langle m | M^*|i \rangle ,
\end{equation}
the right-hand side is zero if there are no loops -- there are no intermediate 
states $|m \rangle$. The right-hand side is also zero 
if we neglect the terms quadratic in $M$. So using 
asymmetries in observables for the same (only forward) process does not 
measure T violation when there are no loops. 

An $s$-channel exchange of a narrow single-particle resonance  like the
$Z$ can also give an imaginary part to the amplitude proportional to the
width of the resonance when it is on shell. However, the width is
equivalent  to the imaginary part of the vacuum polarization loop for
the resonance, and is essentially a loop contribution.

When the
imaginary part of a transition amplitude is absent or negligible, 
the amplitudes
for forward and backward processes are complex conjugates of each other
and therefore equal in magnitude. From unitarity of the $S$ matrix,      this can be seen to be the case when only Born amplitudes are used and
higher-order corrections are absent or neglected. In diagrammatic
language, this amounts to absence of loop corrections.

This has important consequences because of the CPT theorem, which states
that in Lorentz-invariant local field theories, there is invariance
under the combined transformation of C, P and T. However, for the
invariance to hold, T has to be genuine time reversal.
If, instead, a combination
of C, P and naive time reversal T is used, the theorem would
hold if the absorptive part is absent, as reasoned above,
and any violation of CPT in
a process would be proportional to the absorptive part of the amplitude
of the process.

It follows from the above that for a CP odd observable to be nonzero,
the observable has to be T odd in the absence of absorptive
parts. Any T-even contribution to such an observable has to come
from absorptive parts. This will hold mutatis mutandis for a CP-even
observable.

\section{\boldmath Anomalous $HZZ$ vertices in $HZ$ production}

We consider the process $e^+e^-\rightarrow Z H$,
where  the vertex $Z^\ast_{\mu}(k_{1})\rightarrow Z_{\nu}(k_{2}) H$  
has the Lorentz structure
\begin{equation}\label{vertex}
 \Gamma^{V}_{\mu \nu} =\frac{g}{\cos\theta_{W}}m_{Z} \left[ a_{Z}g_{\mu 
\nu}+
\frac{ b_{Z}}{m_{Z}^{2}}\left( k_{1 \nu}k_{2 \mu}-g_{\mu \nu}  k_{1}. 
k_{2}\right) +\frac{\tilde b_{Z}}{m_{Z}^{2}}
\epsilon_{\mu \nu \alpha \beta} k_{1}^{\alpha} k_{2}^{\beta}\right]  ,
 \end{equation}
where $g$ is the $SU(2)_L$ coupling and $\theta_{W}$ is the weak mixing 
angle. The couplings $ a_{Z}$, $b_Z$ and $\tilde b_{Z}$  are 
Lorentz scalars, and
depending on the framework employed, are either real constants, or
complex, momentum-dependent, form factors. The $a_Z$ and $b_Z$ terms 
are invariant under  CP, while  the 
$\tilde b_{Z}$ term corresponds to CP violation. In the SM, at 
tree level, the coupling $ a_{Z}=1$, whereas the other two couplings 
$b_{Z}$ and $\tilde b_{Z}$ vanish.

\subsection{\boldmath Helicity amplitudes for $e^+e^- \to  HZ$}

To calculate these amplitudes we adopt the following representations for the 
transverse and longitudinal polarization vectors of the $Z$:
 \begin{equation}
 \epsilon^{\mu}(k_Z,\pm)=\frac{1}{\sqrt{2}}(0,\mp \cos\theta, - i ,\pm \sin\theta),
  \end{equation}
   \begin{equation}
   \epsilon^{\mu}(k_Z,0)=\frac{1}{m_{Z}}(\vert \vec{k}_{Z}\vert, E_{Z}\sin\theta,0,E_{Z}\cos\theta),
 \end{equation}
 where $E_{Z}$ and $\vec{k}_{Z}$ are the energy and momentum of the $Z$ 
respectively, with $\theta$ being the polar angle made by the $Z$ with respect to the $e^{-}$ momentum taken to be  along the positive $z$ axis.

The non-zero helicity amplitudes in the limit of massless initial 
states are \cite{Rao:2019hsp}
\begin{eqnarray}\label{helamp1}
M(-,+,+)&=&\frac{g^{2}m_{Z}\sqrt{{s}}(c_{V}+c_{A})}
{2\sqrt{2}\cos^{2}\theta_{W}({s}-m_{Z}^{2})}\left[a_Z- \frac{\sqrt{{s}}}{ m_{Z}^{2}}(E_{Z}b_{Z}+ i  \tilde b_{Z}\vert \vec{p}_{Z}\vert)
\right]\\ \nonumber
&& \times  (1-\cos\theta)\\\label{helamp2}
M(-,+,-)&=&\frac{g^{2}m_{Z}\sqrt{{s}}(c_{V}+c_{A})}
{2\sqrt{2}\cos^{2}\theta_{W}({s}-m_{Z}^{2})}\left[a_Z- \frac{\sqrt{{s}}}{ m_{Z}^{2}}(E_{Z}b_{Z}- i  \tilde b_{Z}\vert \vec{p}_{Z}\vert)
\right] \\ \nonumber
&& \times (1+\cos\theta)\\\label{helamp3}
M(-,+,0)&=&\frac{g^{2}E_Z\sqrt{{s}}(c_{V}+c_{A})}
{2\cos^{2}\theta_{W}({s}-m_{Z}^{2})}\left[a_Z-\frac{\sqrt{{s}}}{E_Z}
b_{Z}\right]  \sin\theta \\\label{helamp4}
M(+,-,+)&=&\frac{-g^{2}m_{Z}\sqrt{{s}}(c_{V}-c_{A})}
{2\sqrt{2}\cos^{2}\theta_{W}({s}-m_{Z}^{2})}\left[  a_Z-
  \frac{\sqrt{{s}}}{ m_{Z}^{2}}(E_{Z}b_{Z}+ i  \tilde b_{Z} \vert \vec{p}_{Z}\vert)\right] \\ \nonumber 
  && \times (1+\cos\theta)\\\label{helamp5}
M(+,-,-)&=&\frac{-g^{2}m_{Z}\sqrt{{s}}(c_{V}-c_{A})}
{2\sqrt{2}\cos^{2}\theta_{W}({s}-m_{Z}^{2})}\left[  a_Z-
  \frac{\sqrt{{s}}}{ m_{Z}^{2}}(E_{Z}b_{Z}- i  \tilde b_{Z}\vert \vec{p}_{Z}\vert)\right] \\ \nonumber
  && \times (1-\cos\theta)\\\label{helamp6}
M(+,-,0)&=&\frac{g^{2}E_Z\sqrt{{s}}(c_{V}-c_{A})}{
2\cos^{2}\theta_{W}({s}-m_{Z}^{2})}\left[a_Z-\frac{\sqrt{{s}}}{E_Z}
b_{Z}\right]  \sin\theta
\end{eqnarray}
Here the first two entries in $M$ denote the signs of the helicities of 
the $e^-$ and $e^+$, respectively, and the third entry is the 
$Z$ helicity. $\theta$ is the polar angle of the $Z$ relative to the
$e^-$ direction as the $z$ axis. $c_V$ and $c_A$ are respectively the
vector and axial-vector leptonic couplings of the $Z$, given by
\begin{equation}
c_V = \frac{1}{2}(-1+4\sin^2\theta_W),\;\;
c_A = -\frac{1}{2}.
\end{equation}

\subsection{\boldmath Density matrix for $Z$}

The spin-density matrix for $Z$ production expressed in terms of the helicity amplitudes is given by
 \begin{equation}\label{rhodef}
\rho(i,j)=\overline{ \displaystyle\sum_{\lambda,\lambda^{'}}}
M(\lambda,\lambda^{'},i)M^{\ast}(\lambda,\lambda^{'},j),
 \end{equation}
 the average being over the initial helicities $\lambda$, $\lambda^{'}$ 
of the electron and positron respectively and the indices $i,j$ can take values $\pm,0$.
The diagonal elements of  eq. (\ref{rhodef}) with $i=j$ would give the production probabilities of 
$Z$ with definite polarization, whose ratios to the total cross section are 
known as helicity fractions for the corresponding polarizations. 
Apart from  these  diagonal elements,
it is also necessary to know the off-diagonal elements 
to include the spin information in a coherent way in the combination of the 
$Z$ production and decay processes.  
Then, on integrating over the  phase 
space, the full density matrix eq. (\ref{rhodef}) would lead to the eight independent 
vector and tensor polarization components, known as the 
\textit{polarization parameters} of the $Z$. 

When the beams are polarized the expression we need to use is
 \begin{equation}\label{rhobeampol}
\displaystyle \rho(i,j)=\frac{1}{4}
\sum_{\lambda,\lambda^{'}} (1+\lambda P_L)(1+\lambda' \bar P_L)
M(\lambda,\lambda^{'},i)M^{\ast}(\lambda,\lambda^{'},j),
 \end{equation}
which is the same as eq. (\ref{denmatpol}), without the index $h$,
since the Higgs in unpolarized.

The density matrix elements for unpolarized $e^+$ and $e^-$ beams 
derived from the helicity amplitudes 
are given by
\begin{eqnarray}
 \rho(\pm,\pm
)&=&\frac{g^{4}m^{2}_{Z}s}{8\cos^{4}\theta_{W}(s-m_{Z}^{2})^2}
\left[
(c_{V}+c_{A})^{2}(1\mp\cos\theta)^{2}\right. \nonumber\\
&& \hskip -0.8cm \left. +(c_{V}-c_{A})^{2}(1\pm\cos\theta)^{2}\right]
\left| a_Z 
-(b_{Z}\pm i\beta_{Z}\tilde b_{Z})\frac{
E_{Z}\sqrt{s}}{m^{2}_{Z}} \right|^2 \\
\rho(0,0
)&=&\frac{g^{4}E^{2}_{Z}s}{2\cos^{4}\theta_{W}(s-m_{Z}^{2})^2}\sin^{2}\theta
\, (c_V^2 + c_A^2) 
\left| a_Z -b_{Z}\frac{\sqrt{s}}{E_{Z}}\right|^2  \\
\rho(\pm,\mp
)&=&\frac{g^{4}m^{2}_{Z}s}{4\cos^{4}\theta_{W}(s-m_{Z}^{2})^2}\sin^{2}\theta
 \,(c_V^2 + c_A^2)\nonumber \\ &&\times 
\left[\left| a_Z  - b_{Z}\frac{\sqrt{s}E_Z}{m_Z^2} \right|^2 \mp  
2i {\rm Re} \left((a_Z
- \frac{\sqrt{s}E_Z}{m_Z^2}b_Z)  \tilde b_{Z}^*\right)\frac{
E_{Z}\beta_Z\sqrt{s}}{m^{2}_{Z}} \right. \nonumber \\
&& \left. - \frac{s\beta_Z^2 E_Z^2 }{m_Z^4}
\vert \tilde b_Z \vert^2 \right]  \\
\rho(\pm,0 )&=&\frac{g^{4}m_{Z}E_{Z}s}
{4\sqrt{2}\cos^{4}\theta_{W}(s-m_{Z}^{2})^2}\sin\theta
\nonumber \\
&& \hskip -.8cm \times \left[
(c_{V}+c_{A})^{2}(1\mp\cos\theta) 
 -(c_{V}-c_{A})^{2}(1\pm\cos\theta)\right] \nonumber\\
&& \hskip -.8cm \times 
\left[\vert a_Z \vert^2-{\rm Re}(a_Zb_{Z}^*)\frac{\sqrt{s}
(E^{2}_{Z}+m^{2}_{Z})}{E_{Z}m^{2}_{Z}}
 + i {\rm Im}(a_Zb^*_Z)\frac{\sqrt{s} 
E_{Z}\beta_Z^2}{m^{2}_{Z}} \right. \nonumber\\
&& \left. + \frac{s}{m_Z^2}\vert b_Z\vert^2 
\mp i  \tilde b_Z \frac{\sqrt{s}E_Z\beta_Z}{m_Z^2}
(a_Z^* - b_{Z}^* \frac{\sqrt{s}}{E_Z})\right] 
\end{eqnarray}
where $\beta_{Z}=\vert\vec{k}_{Z}\vert /E_{Z}$ is the velocity of the
$Z$ in the c.m. frame. 
We do not display the somewhat longer expressions for the density matrix
 elements taking into account the polarizations $P_L$ and $\bar{P}_L$ of
 the electron and positron beams, respectively. 
However, the expressions are 
more 
compact on integration over $\cos\theta$, and these are displayed here.
To carry out this integration, 
we include the appropriate phase space factor, which is the same as 
the one relating
the differential cross section to the 
square of the matrix element $M$ in the expression:
\begin{equation}\label{phasespace}
\frac{d\sigma}{d\cos\theta} = \sum_{\lambda,\lambda'}\vert
M \vert^2 \frac{\vert \vec k_Z \vert }{64 \pi s^{3/2}}.
\end{equation}
The
expressions are then normalized to give the correct total
cross section $\sigma$ as the trace of the density matrix as given in
eq. (\ref{totalcs}).

The integrated density matrix is given by
\begin{eqnarray}\label{sigmapmpm}
 \sigma(\pm,\pm
)&=&\frac{2(1-P_L\bar P_L)g^{4}m^{2}_{Z}\vert \vec k_Z \vert}{192\pi
\sqrt{s} \cos^{4}\theta_{W}(s-m_{Z}^{2})^2}
(c_{V}^2+c_{A}^{2}-2P_L^{\rm eff}c_Vc_A) \nonumber\\
&& \hskip -0.8cm 
\times
\left| a_Z 
-(b_{Z}\pm i\beta_{Z}\tilde b_{Z})\frac{
E_{Z}\sqrt{s}}{m^{2}_{Z}} \right|^2 \\ \label{sigma00}
\sigma(0,0
)&=&\frac{2(1-P_L\bar P_L)g^{4}E^{2}_{Z}\vert\vec k_Z \vert}
{192\pi\sqrt{s}\cos^{4}\theta_{W}(s-m_{Z}^{2})^2}
(c_V^2 + c_A^2 -2P_L^{\rm eff}c_Vc_A)\nonumber \\ 
&& \hskip -0.8cm
\times \left| a_Z -b_{Z}\frac{\sqrt{s}}{E_{Z}}\right|^2  \\
\label{sigmapmmp}
\sigma(\pm,\mp)&=&\frac{(1-P_L\bar P_L)g^{4}
m^{2}_{Z}\vert \vec k_Z \vert }
{192\pi\sqrt{s}\cos^{4}\theta_{W}(s-m_{Z}^{2})^2}
 \,(c_V^2 + c_A^2 - 2P_L^{\rm eff}c_Vc_A)\nonumber \\ 
&& \hskip -0.8cm
\times\left[\left| a_Z  - b_{Z}\frac{\sqrt{s}E_Z}{m_Z^2} \right|^2 \mp  
2i {\rm Re} \left((a_Z
- \frac{\sqrt{s}E_Z}{m_Z^2}b_Z)  \tilde b_{Z}^*\right)\frac{
\vert\vec k_Z\vert\sqrt{s}}{m^{2}_{Z}} \right. \nonumber \\
&& \left. - \frac{s \vert \vec k_Z \vert^2 }{m_Z^4}
\vert \tilde b_Z \vert^2 \right]  \\\label{sigmapm0}
\sigma(\pm,0 )&=&\frac{(1-P_L\bar P_L)g^{4}m_{Z}E_{Z}\vert\vec k_Z\vert}
{256\sqrt{2}\sqrt{s}\cos^{4}\theta_{W}(s-m_{Z}^{2})^2}
 (2c_{V}c_{A} - P_L^{\rm eff}(c_V^2+c_A^2))
\nonumber\\
&& \hskip -.8cm 
\times \left[\vert a_Z \vert^2-{\rm Re}(a_Zb_{Z}^*)\frac{\sqrt{s}
(E^{2}_{Z}+m^{2}_{Z})}{E_{Z}m^{2}_{Z}}
 + i {\rm Im}(a_Zb^*_Z)\frac{\sqrt{s} 
E_{Z}\beta_Z^2}{m^{2}_{Z}} \right. \nonumber\\
&& \left. + \frac{s}{m_Z^2}\vert b_Z\vert^2 
\mp i  \tilde b_Z \frac{\vert\vec k_Z\vert\sqrt{s}}{m_Z^2}
(a_Z^* - b_{Z}^* \frac{\sqrt{s}}{E_Z})\right].
\end{eqnarray}
\nopagebreak\\
In the above equations, $P_L^{\rm eff} = (P_L - \bar P_L)/ 
( 1 - P_L \bar P_L)$, and the indices $+$, $-$ and $0$ denote the $Z$
helicities.

These expressions for the integrated density matrix elements will be
used later to determine the angular distributions and hence the angular
asymmetries of leptons produced in
$Z$ decay. In addition, we will also discuss a combination of the lepton
angular asymmetries combined with the forward-backward asymmetry of the
$Z$. To this end, we need to list the density matrix elements
antisymmetrized over the forward and backward hemispheres. In other
words, the integral over the backward hemisphere ($\pi/2 < \theta <
\pi$) is subtracted from the integral over the forward hemisphere ($0 <
\theta < \pi/2$). The corresponding expressions for the antisymmetrized
density matrix elements, denoted with a prime, are as follows:
\begin{eqnarray}\label{sigmap}
 \sigma'(\pm,\pm
)&=&\mp\frac{(1-P_L\bar P_L)g^{4}m^{2}_{Z}\vert \vec k_Z \vert}{128\pi
\sqrt{s} \cos^{4}\theta_{W}(s-m_{Z}^{2})^2}
(2c_Vc_A-(c_{V}^2+c_{A}^{2})P_L^{\rm eff}) \nonumber\\
&& \hskip -0.8cm 
\times
\left| a_Z 
-(b_{Z}\pm i\beta_{Z}\tilde b_{Z})\frac{
E_{Z}\sqrt{s}}{m^{2}_{Z}} \right|^2 \\
\sigma'(0,0)&=&0\\
\sigma'(\pm,\mp)&=&0\\
\sigma'(\pm,0 )&=&\mp\frac{(1-P_L\bar P_L)g^{4}m_{Z}E_{Z}\vert\vec k_Z
\vert}
{192\pi\sqrt{2}\sqrt{s}\cos^{4}\theta_{W}(s-m_{Z}^{2})^2}
(c_V^2+c_A^2 - 2 P_L^{\rm eff} c_V c_A)
\nonumber\\
&& \hskip -.8cm 
\times \left[\vert a_Z \vert^2-{\rm Re}(a_Zb_{Z}^*)\frac{\sqrt{s}
(E^{2}_{Z}+m^{2}_{Z})}{E_{Z}m^{2}_{Z}}
 + i {\rm Im}(a_Zb^*_Z)\frac{\sqrt{s} 
E_{Z}\beta_Z^2}{m^{2}_{Z}} \right. \nonumber\\
&& \left. + \frac{s}{m_Z^2}\vert b_Z\vert^2 
\mp i  \tilde b_Z \frac{\vert\vec k_Z\vert\sqrt{s}}{m_Z^2}
(a_Z^* - b_{Z}^* \frac{\sqrt{s}}{E_Z})\right].
\end{eqnarray}
\nopagebreak\\

\subsection{Explicit expressions for the asymmetries}
We now present explicit expressions for the angular asymmetries of the
charged lepton $\ell^-$ in the process
$e^+e^- \to HZ$, followed by $Z \to \ell^+ \ell^-$. These are  
obtained by substituting the expressions for the density matrix elements
eqs. (\ref{sigmapmpm})-(\ref{sigmapm0}) into eqs. (\ref{pol1})-(\ref{pol9}) and using eqs.
(\ref{asyx})-(\ref{asyx2y2}).
 \begin{eqnarray}\label{asy1}
A_{x}&=&  
\frac{(1-P_L\bar P_L)3\alpha g^{4}m_{Z}E_{Z}\vert\vec k_Z\vert}
{512\sigma\sqrt{s}\cos^{4}\theta_{W}(s-m_{Z}^{2})^2}
 (2c_{V}c_{A} - P_L^{\rm eff}(c_V^2+c_A^2))
\nonumber\\
&& \hskip -.8cm 
\times \left[\vert a_Z \vert^2-{\rm Re}(a_Zb_{Z}^*)\frac{\sqrt{s}
(E^{2}_{Z}+m^{2}_{Z})}{E_{Z}m^{2}_{Z}}
 + \frac{s}{m_Z^2}\vert b_Z\vert^2 
\right]
 .\\\label{asy2}
 A_{y}&=&-
\frac{(1-P_L\bar P_L)3\alpha g^{4}m_{Z}E_{Z}\vert\vec k_Z\vert}
{512\sigma\sqrt{s}\cos^{4}\theta_{W}(s-m_{Z}^{2})^2}
 (2c_{V}c_{A} - P_L^{\rm eff}(c_V^2+c_A^2))
\nonumber\\
&& \hskip -.8cm 
\times 
{\rm Re} \left( 
(a_Z - b_{Z} \frac{\sqrt{s}}{E_Z})\tilde b^*_Z\right)
\frac{\vert\vec k_Z\vert\sqrt{s}}{m_Z^2}.
\\\label{asy3}
 A_{z}&=&
-\frac{2(1-P_L\bar P_L)\alpha g^{4}m^{2}_{Z}\vert \vec k_Z \vert}{64\pi\sigma
\sqrt{s} \cos^{4}\theta_{W}(s-m_{Z}^{2})^2}
(c_{V}^2+c_{A}^{2}-2P_L^{\rm eff}c_Vc_A) \nonumber\\
&& \hskip -0.8cm 
\times
 {\rm Im}\left((a_Z -b_{Z} \frac{ \sqrt{s}E_Z}{m^{2}_{Z}}) 
\tilde b^*_{Z}\right)
\frac{
\vert\vec k_{Z}\vert\sqrt{s}}{m^{2}_{Z}} \\
 \label{asy4}
 A_{xy}&=&
-\frac{(1-P_L\bar P_L)g^{4}
m^{2}_{Z}\vert \vec k_Z \vert }
{48\pi^2\sigma\sqrt{s}\cos^{4}\theta_{W}(s-m_{Z}^{2})^2}
 \,(c_V^2 + c_A^2 - 2P_L^{\rm eff}c_Vc_A)\nonumber \\ 
&& \hskip -0.8cm
\times   
 {\rm Re} \left((a_Z
- \frac{\sqrt{s}E_Z}{m_Z^2}b_Z) \tilde b_{Z}^*\right) \frac{
\vert\vec k_{Z}\vert\sqrt{s}}{m^{2}_{Z}}   \\
\label{asy5}
  A_{xz}&=& 
\frac{(1-P_L\bar P_L)g^{4}m_{Z}E_{Z}\vert\vec k_Z\vert}
{128\pi\sigma\sqrt{s}\cos^{4}\theta_{W}(s-m_{Z}^{2})^2}
 (2c_{V}c_{A} - P_L^{\rm eff}(c_V^2+c_A^2))
\nonumber\\
&& \hskip -.8cm 
\times 
  {\rm Im} \left( 
(a_Z - b_{Z} \frac{\sqrt{s}}{E_Z})
\tilde b^*_Z \right) \frac{\vert\vec k_Z\vert\sqrt{s}}{m_Z^2}).\\\label{asy6}
  A_{yz}&=&
-\frac{(1-P_L\bar P_L)g^{4}m_{Z}E_{Z}\vert\vec k_Z\vert}
{128\pi\sigma\sqrt{s}\cos^{4}\theta_{W}(s-m_{Z}^{2})^2}
 (2c_{V}c_{A} - P_L^{\rm eff}(c_V^2+c_A^2))
\nonumber\\
&& \hskip -.8cm 
\times 
 {\rm Im}(a_Zb^*_Z)\frac{\sqrt{s} 
E_{Z}\beta_Z^2}{m^{2}_{Z}} . \\ \label{asy7}
  A_{x^2-y^2}&=&
\frac{(1-P_L\bar P_L)g^{4}
m^{2}_{Z}\vert \vec k_Z \vert }
{96\pi^2\sigma\sqrt{s}\cos^{4}\theta_{W}(s-m_{Z}^{2})^2}
 \,(c_V^2 + c_A^2 - 2P_L^{\rm eff}c_Vc_A)\nonumber \\ 
&& \hskip -0.8cm
\times\left[\left| a_Z  - b_{Z}\frac{\sqrt{s}E_Z}{m_Z^2} \right|^2 
 - \frac{s \vert \vec k_Z \vert^2 }{m_Z^4}
\vert \tilde b_Z \vert^2 \right]  \\ \label{asy8}
  A_{zz}&=&
 \frac{9}{16}\left[\frac{1}{3}-
\frac{2(1-P_L\bar P_L)g^{4}E^{2}_{Z}\vert\vec k_Z\vert}
{192\pi\sigma\sqrt{s}\cos^{4}\theta_{W}(s-m_{Z}^{2})^2}
(c_V^2 + c_A^2 -2P_L^{\rm eff}c_Vc_A)\right.\nonumber \\ 
&& \hskip -0.8cm
\left. \times \left| a_Z -b_{Z}\frac{\sqrt{s}}{E_{Z}}\right|^2  
\right].
 \end{eqnarray}

We can now check the entries in Table 1 against these expressions for the
asymmetries. The first important prediction from Table 1 was that the
asymmetries $A_y$, $A_z$, $A_{xy}$ and $A_{xz}$ are odd under CP. We see 
from eqs. (\ref{asy2}), (\ref{asy3}), (\ref{asy4}) and (\ref{asy5})
 that these asymmetries are
proportional to $\tilde b_Z$, which is a CP-violating coupling.
Further, Table 1 shows that the asymmetries $A_z$, $A_{xz}$ and $A_{yz}$ are
odd under CPT. We see in the above equations that it is precisely these
asymmetries which are proportional to the imaginary part of the product
of couplings, required by consistency with the CPT theorem.

We now look at the asymmetries $A'_i$ which are the asymmetries $A_i$
combined with the forward-backward asymmetry of the produced $Z$ boson
in the laboratory frame. Expressions for these are easily written down.
These are obtained from 
eqs. (\ref{asyx})-(\ref{asyzz}) by replacing the various
$\sigma(i,j)$ in the polarization vectors and tensors in the last member
of the equations by the respective $\sigma'(i,j)$.
 \begin{eqnarray}\label{pol'1}
A'_{x}&=&  
\frac{(1-P_L\bar P_L)\alpha g^{4}m_{Z}E_{Z}\vert\vec k_Z\vert}
{128\pi\sigma\sqrt{s}\cos^{4}\theta_{W}(s-m_{Z}^{2})^2}
(c_V^2+c_A^2 - 2 P_L^{\rm eff} c_V c_A)
\nonumber\\
&& \hskip -.8cm 
\times 
{\rm Im}\left( 
(a_Z - b_{Z} \frac{\sqrt{s}}{E_Z})
\tilde b^*_Z \right)\frac{\vert\vec k_Z\vert\sqrt{s}}{m_Z^2}.\\ \label{pol'2}
A'_{y}&=&
\frac{(1-P_L\bar P_L)\alpha g^{4}m_{Z}E_{Z}\vert\vec k_Z\vert}
{128\pi\sigma\sqrt{s}\cos^{4}\theta_{W}(s-m_{Z}^{2})^2}
(c_V^2+c_A^2 - 2 P_L^{\rm eff} c_V c_A)
\nonumber\\
&& \hskip -.8cm 
\times {\rm Im}(a_Zb_{Z}^*)\frac{\sqrt{s}
E_{Z}\beta_Z^2}{m^{2}_{Z}}.\\ \label{pol'3}
 A'_{z}&=&
-\frac{(1-P_L\bar P_L)3\alpha g^{4}m^{2}_{Z}\vert \vec k_Z \vert}{512\pi
\sigma\sqrt{s} \cos^{4}\theta_{W}(s-m_{Z}^{2})^2}
(2c_Vc_A-(c_{V}^2+c_{A}^{2})P_L^{\rm eff}) \nonumber\\
&& \hskip -0.8cm 
\times
\left[
\left| a_Z 
- b_Z \frac{E_Z\sqrt{s}}{m_Z^2} \right|^2 
+ \left| \tilde b_Z\frac{\vert\vec k_Z\vert\sqrt{s}}{m^2_Z} \right|^2 \right] 
\\ \label{pol'4}
 A'_{xy}&=&0\\\label{pol'5}
  A'_{xz}&=&
\frac{(1-P_L\bar P_L)g^{4}m_{Z}E_{Z}\vert\vec k_Z\vert}
{96\pi^2\sigma\sqrt{s}\cos^{4}\theta_{W}(s-m_{Z}^{2})^2}
(c_V^2+c_A^2 - 2 P_L^{\rm eff} c_V c_A)
\nonumber\\
&& \hskip -.8cm 
\times \left[\vert a_Z \vert^2-{\rm Re}(a_Zb_{Z}^*)\frac{\sqrt{s}
(E^{2}_{Z}+m^{2}_{Z})}{E_{Z}m^{2}_{Z}} 
+ \frac{s}{m_Z^2}\vert b_Z\vert^2 \right]\\\label{pol'6}
  A'_{yz}&=&-
\frac{(1-P_L\bar P_L)g^{4}m_{Z}E_{Z}\vert\vec k_Z\vert}
{96\pi^2\sigma\sqrt{s}\cos^{4}\theta_{W}(s-m_{Z}^{2})^2}
(c_V^2+c_A^2 - 2 P_L^{\rm eff} c_V c_A)
\nonumber\\
&& \hskip -.8cm 
\times 
{\rm Re} \left( 
 (a_Z - b_{Z} \frac{\sqrt{s}}{E_Z})
 \tilde b^*_Z \right)\frac{\vert\vec k_Z\vert\sqrt{s}}{m_Z^2}
).\\\label{pol'7}
  A'_{x^2-y^2}&=&0\\\label{pol'8}
  A'_{zz}& = & 
-\frac{(1-P_L\bar P_L)g^{4}m^{2}_{Z}\vert \vec k_Z \vert}{64\pi
\sigma\sqrt{s} \cos^{4}\theta_{W}(s-m_{Z}^{2})^2}
(2c_Vc_A-(c_{V}^2+c_{A}^{2})P_L^{\rm eff}) \nonumber\\
&& \hskip -0.8cm 
\times
 {\rm Im}\left( (a_Z 
-b_{Z}
\frac{
E_{Z}\sqrt{s}}{m^{2}_{Z}}) 
\tilde b^*_{Z}\right)
\frac{
\vert\vec k_{Z}\vert\sqrt{s}}{m^{2}_{Z}}.
 \end{eqnarray}
We find that 
$A'_{xy}$ and $A'_{x^2-y^2}$ are vanishing
regardless of the couplings.

We can verify that the asymmetries which are shown to be CP odd in Table
 2, namely, $A'_x$,  $A'_{yz}$ and $A'_{zz}$
are proportional to $\tilde b_Z$ in the above equations.
Table 2 also shows $A'_{x^2-y^2}$,  
as CP odd. However, for
our process, $A'_{x^2-y^2}$ turns out to be vanishing
independent of the couplings.
 As for
CPT, $A'_x$, $A'_y$ and $A'_{zz}$ are proportional to the imaginary parts of
products of couplings, consistent with their being CPT odd as seen in
Table 2. Table 2 also shows $A'_{xy}$ and $A'_{x^2-y^2}$ as CPT odd, 
but they turn out to be zero for our process.

\subsection{When does beam polarization help? An example}

That a combination of opposite-sign beam polarizations for the $e^-$ and
$e^+$ enhances certain asymmetries can
be seen from the following argument. This argument applies only to processes
which are mediated by an $s$-channel exchange of a virtual photon or a
virtual $Z$. We illustrate it for the specific process $e^+e^- \to HZ$.

Consider an asymmetry which is odd under P
and also odd under T, as for example $A_{yz}$.
Now, all momenta transform in the same way under 
P as under T, i.e., they reverse their signs. In addition, under P, the
helicities change sign. Thus, if P is a symmetry of the theory, $A_{yz}$
being odd under P would vanish. Our theory is not symmetric under P,
since the left-hand and right-handed couplings of $Z$ to leptons,
respectively, $L_\ell$ and $R_\ell$ are
different. In cases when the process has an $s$-channel exchange of a
virtual $Z$, the $e^+e^-$ pair couples to the $Z$ through a vertex with
a mix of left-handed and right-handed chiral couplings.
Because the difference $R_\ell^2 - L_\ell^2$,
which is proportional to  $c_V c_A \approx -0.03$,  
is numerically small, there is an approximate P
invariance. The asymmetry turns out to be proportional to $c_V$ and
is therefore small. In the presence of significant 
beam polarizations which are 
opposite in sign for the $e^-$ and $e^+$, the left-handed and
right-handed couplings are not longer approximately equal, and the 
P symmetry is no longer an
approximate symmetry. The small quantity $c_Vc_A$ gets replaced, in the
presence of electron and positron beam polarizations $P_L$ and
$\bar P_L$ by $(c_Vc_A - \frac{1}{2} (c_V^2 + c_A^2) P_L^{\rm eff})$, where 
\begin{equation}\label{pleff}
P_L^{\rm eff} = (P_L - \bar P_L)/ 
( 1 - P_L \bar P_L),
\end{equation}
This can be
quite large, depending on the value of the beam polarizations, and the
parity violation can be large.  As a
result, $A_{yz}$ is greatly enhanced.

We can see how this works out in practice from explicit expressions.
$A_{yz}$ arises from a combination of the imaginary parts of the 
density matrix elements $\rho(\pm,0)$. It can be seen from eqs.
(\ref{helamp1})-(\ref{helamp6})
that the contribution to these matrix elements from the $e^-$ and $e^+$
helicity combinations $(-,+)$ and $(+,-)$ occur with opposite signs,
in addition to the different couplings $(c_V+c_A)$ and $(c_V-c_A)$,
respectively. Since numerically $c_V$ is much smaller than $c_A$ in
magnitude, this results in a partial cancellation in the calculation of
$\rho(+,0)$ as well as that of $\rho(-,0)$, giving a coupling
dependence of $2c_Vc_A$ in the numerator of the asymmetry as compared 
to $c_V^2+c_A^2$ in the  
cross section appearing in the denominator of the asymmetry.
In the presence of polarization, the factors $(c_V+c_A)^2$ and 
$(c_V-c_A)^2$ get different polarization dependent factors, preventing
the partial cancellation, and thus enhancing the asymmetry.

This was an example of an asymmetry which is expected to be small in
the unpolarized case, but is enhanced by beam polarization. Let us
examine which other symmetries are expected to be larger with beam
polarization. 

\subsection{Which asymmetries get enhanced with beam polarization?}

In our process, there are three possible sources of P violation -- P
violation at the $e^+e^-Z$ vertex, P violation at the $Z\ell^+\ell^-$ 
decay vertex, and P violation at the $HZZ$ vertex.
With unpolarized beams, the asymmetries which will be
suppressed because of approximate P symmetry 
are those which have parity violation at the $e^+e^-Z$ 
vertex. We need to single out these asymmetries. 

To find out which asymmetries are enhanced by polarization, we need t
o find asymmetries which contain parity violation at the $e^+e^-Z$ 
vertex, whether or not they violate parity at the other two ($HZZ$ and 
$Z\ell^+\ell^-)$ vertices.
From the definition of the asymmetries, we only know whether 
they violate parity overall or not. This information is in Table 1.  From Table 1, 
we know which asymmetries violate parity (overall). We would like to find 
out, by elimination, which asymmetries 
 violate parity at the $e^+e^-Z$ vertex,
whether they violate parity overall or not.

These are the two possibilities, (A) and (B), corresponding to whether
there is overall parity
 violation (A) or not (B). Writing the parity properties
 of the $e^+e^-Z$, $HHZ$ and $Z\ell^+\ell^-$ vertices as odd (O) or 
even (E), 
we can characterize the (A) category as one of (OOO), (OEE), (EOE) or 
(EEO).
Here, the order of the 3 letters is chosen as representing the parity
property of respectively the $e^+e^-Z$ vertex, the $HHZ$ vertex and the 
$Z\ell^+\ell^-$ vertex.  
 Similarly the (B) category has one of (EEE), (OOE), (OEO), or (EOO). 

 Now CP violation can come only from the $HZZ$ vertex. So if the asymmetry
 is overall CP violating, then the $HZZ$ vertex has a Levi-Civita factor, 
which also violates parity. So the CP violating asymmetries from Table 1
 will have O in the middle place.

The asymmetries violating P at the $Z\ell^+\ell^-$ vertex have a factor 
of 
$\alpha$ proportional to $c_Vc_A$ in the asymmetry (see eq.
\ref{polanalyser}). These are $A_x$,
$A_y$ and $A_z$, which will all have O in the third place.

So to pick out asymmetries which have O in the first place, 
by elimination, from (A) we should select those which have OO in the 
second and third places, or EE in the second and third places. O in the
second place means CP violating, and O in the third place means one of
$A_x$, $A_y$ and $A_z$. This is the category (A)(1). The category (A)(2)
 is with 
E in the second place (CP conserving) and E in the third place (not one 
of $A_x$, $A_y$, $A_z$).

Similarly, from (B) we should select EO or OE in the second and third 
places. These are, respectively, CP conserving and one of $A_x$, $A_y$, 
$A_z$ (B)(1), and CP violating and not one of $A_x$, $A_y$, $A_z$ (B)(2).

\comment{
Asymmetries which violate P at the $e^+e^-Z$ vertex are 
isolated by looking at asymmetries with overall P violation, 
but with either
no P violation at either the $HZZ$ vertex or the $Z$ decay vertex, or P
violation at both $HZZ$ and $Z$ decay vertices. Alternatively, we look 
for
asymmetries with no overall P violation, but P violation at either one 
of the
$HZZ$ and $Z$ decay vertices, so that this P violation is compensated by
P violation at the $e^+e^-Z$ vertex.
}

The final result is that asymmetries which violate P at the $e^+e^-Z$ 
vertex are
\begin{itemize}
\item[(A)] P violating asymmetries but 
\begin{itemize} \item[(1)] CP violating asymmetries out of
 $A_x$, $A_y$ and $A_z$,  or,  
\item[(2)] CP conserving asymmetries out of
$A_{xy}$, $A_{yz}$ and  $A_{xz}$,
\end{itemize} \end{itemize} 
or, \begin{itemize} \item[(B)] P conserving asymmetries but 
\begin{itemize}
\item[(1)] CP conserving asymmetries out 
of $A_x$, $A_y$ and $A_z$,  or,  
\item[(2)] CP violating asymmetries out
$A_{xy}$, $A_{yz}$ and  $A_{xz}$. \end{itemize}
\end{itemize}

 We conclude that the asymmetries which have P violation at the 
$e^+e^-Z$ vertex are $A_y$, $A_{yz}$, $A_x$ and $A_{xz}$, respectively in 
the categories (A)(1), (A)(2), (B)(1) and  (B)(2).
These would be enhanced with polarized beams. 

At the risk of sounding repetitive, let us take another example, say of 
the category (A)(1) to try and understand better our arguments as to 
 why CP-violating asymmetries out of $A_x$, $A_y$, $A_z$ 
will have P violation at the $e^+e^-Z$ vertex.
In the specific case
of the asymmetry $A_x$, it is  seen from Table 1, that it is overall P and
 CP even. Also it is proportional to $\alpha$, which indicates that there 
is P violation at the $Z\ell^+\ell^-$ vertex. Therefore, to restore the 
P property, there must be another P violation and that can only come
from the $e^+e^-Z$ vertex.

\subsection{\boldmath What about $\tau$ polarization?}

The asymmetries $A_x$, $A_y$ $ A_z$ vanish in the limit of parity 
conservation
in $Z$ decay into fermions. Hence  observing the polarization of the 
fermion ($\tau$) would result in larger asymmetries as compared to the
respective spin-summed asymmetries.

This consideration is independent of the production process. 
So it applies to
all processes $e^+e^- \to Z +X$, and not just to $e^+e^- \to HZ$. 
Here we concentrate on the latter process.

Of these $A_x$ is even under all of C,P,T. Since it needs parity violation
at the $Z$ decay vertex, there must be parity violation at either at 
the $e^+e^-Z$ 
vertex or the $HZZ$ vertex. Since it is CP conserving, the parity 
violation
cannot be at the $HZZ$ vertex, and must be at the $e^+e^-Z$ vertex. 
So $A_x$ is
enhanced by polarized beams, and also by $\tau$ polarization. 

$A_y$ is P violating. So because it is CP violating, and therefore P
violating at the $HZZ$ vertex, it should be P
violating also at the $e^+e^-Z$ vertex. Hence it is enhanced by both beam
polarization and $\tau$ polarization.

$A_z$ is P conserving but CP violating. So it conserves P at the $e^+e^-Z$ 
vertex, and hence not enhanced by beam polarization, but by $\tau$
polarization.

That $A_x$, $A_y$ and $A_z$ vanish in the limit of parity conservation
can be seen from the fact that their expressions are proportional to
$\alpha$, which vanishes if the fermion coupling is either purely vector
or purely axial-vector. When the decay-fermion helicities are not summed
over, the 
dependence of the $Z \to f \bar f$ decay density matrix on the
fermion helicity $h_f$ is of the form $2 c_Vc_A  + 2h_f (c_V^2 + 
c_A^2)$. It is understood that the anti-fermion helicity is $h_{\bar
f} = - h_f$ because of the vector-axial-vector coupling of the fermion
pair to $Z$. Thus, summing over the fermion helicities gives $4 c_V
c_A$, whereas the contribution when $h_f = \pm 1/2$ is $2 c_V c_A \pm
(c_V^2 + c_A^2)$. Because $c_V$ is numerically small, the contribution
when the helicity is measured is much larger than that when helicities
are summed over. Thus, the asymmetries $A_x$, $A_y$ and $A_z$ are
larger when the $\tau$ polarization is nonzero.

\subsection{Beam and $\tau$ polarization dependences of $A'_{i}$}

In going from the $A_i$ asymmetries to the $A'_i$ asymmetries, the
transformation properties
change in the following way:

1. The C properties change, but the P and T properties do not change. 

2. Consequently, the CP properties change.

Though the CP properties change, it is only
because of the CP-odd $Z$ FB asymmetry. This does not change the P
properties. However, the only source of CP-violating interaction is the
$HZZ$
coupling. The change in the CP property  comes from that coupling, which 
contributes a $-1$ factor for P. So asymmetries which are enhanced by 
beam polarization are now exactly those which were not enhanced without
the $Z$ FB asymmetry, and vice versa. 

3. The CPT properties change -- those asymmetries which needed
an absorptive part, now combined with the $Z$ FB asymmetry, 
will not need an absorptive part, and vice versa.

4.  The effect of $\tau$ polarization depends only on the P property of the
$Zf^+f^-$ vertex, so it is the same as in the case of the $A_i$ asymmetries:
enhanced for $A'_x$, $A'_y$, $A'_z$.

%
\section{Measurement of one-loop contribution of the triple-Higgs
coupling}

The SM makes a prediction for the triple-Higgs coupling in terms of the
weak coupling and vacuum expectation value of the neutral scalar field.
This gives a concrete relation between  the triple-Higgs coupling and 
the quartic-Higgs coupling. However, if the SM is an effective field
theory, with BSM contributions appearing at higher energy scales, 
there is
no such simple relation. The triple Higgs coupling can receive
contributions from higher-dimensional operators. An experimental
measurement of the triple-Higgs coupling can be made by studying
two-Higgs production at a hadron collider. However, the corresponding
cross section is significant only at fairly high energies. There have been
proposals to study the triple-Higgs coupling in single-Higgs production
where the triple-Higgs vertex appears at the loop level
\cite{McCullough:2013rea}.  In particular,
at an $e^+e^-$ collider such a loop contribution can occur in the
standard $HZ$ production process, and 
the sensitivity at various
proposals for such colliders has been estimated. The drawback of this
procedure is that in extensions of the SM, even anomalous tree-level
$HZZ$ couplings can give a large contribution, masking the smaller loop
contribution of the $HHH$ coupling. 

One proposal to evade this problem
is to look for observables which are CPT odd \cite{Nakamura:2018bli,
Rao:2021eer}, necessarily requiring the
presence of a loop contribution. Since the triple-Higgs coupling in the
simplest case is CP even, it would contribute to the imaginary
part of the form factor $b_Z$. 
Of the asymmetries we considered in a previous
section, we could look for those asymmetries which are CP
even and CPT odd. Looking at Tables 1 and 2, we can see that the
asymmetries we are looking for are $A_{yz}$, $A'_{xy}$ and $A'_{y}$.
Of these, $A_{yz}$ and $A'_{xy}$ are enhanced by beam polarization.
Only one of these, that is $A'_{y}$, is enhanced by $\tau$ polarization.

In an earlier work \cite{Rao:2021eer},
we studied the possibility  of $A_{yz}$ in 
the measurement of the triple-Higgs coupling. In that work, the lepton
asymmetries combined with $Z$ FB asymmetry were not considered. The
latter were included in the discussion in \cite{Nakamura:2018bli}
but they concluded that
$A'_{xy}$ does not get contribution from diagrams containing a loop 
with a triple-Higgs coupling, and restricted themselves to $A_{yz}$ and
$A'_{y}$. We have however seen that on general grounds, $A'_{xy}$
vanishes for arbitrary $HZZ$ couplings.

\section{Conclusions and Discussion}

We have examined several general expectations in processes of the type
$e^+e^- \to Z + X$, with $Z$ decaying into a pair of charged leptons, with
an emphasis on properties of angular asymmetries of charged leptons
which can be deduced on the basis of discrete transformations C, P, T,
and combinations of these. 
The specific transformations P, CP, T and CPT are of special importance
and we have discussed the consequences of these symmetries or their 
absence for the leptonic asymmetries. 
Importance has also been given to the role of beam
polarization, and to deducing  on general grounds which 
asymmetries get enhanced by the use of beam polarization. Similarly, we
also discuss which asymmetries would be sensitive to the measurement of
$\tau$
polarization in $Z$ decay in to $\tau^+\tau^-$.

We have considered polar and azimuthal asymmetries of decay leptons $A_i$, 
as well
as combinations $A'_i$ of the asymmetries $A_i$ with a $Z$ forward-backward 
asymmetry. These two sets of asymmetries are sensitive to different sets of
couplings.
 
Using a general set of arguments to determine the CP and CPT
properties of lepton angular asymmetries we can thus choose asymmetries
which would be CP even or CP odd, as also CPT even or CPT odd. The
latter property determines whether the real part or the imaginary part
of the combination of couplings contributes.
For a given combination of coupling constants -- whether CP even or CP
odd, and whether the real part of the combination contributes 
or the imaginary part, one
has the option of choosing more than one asymmetry from the sets
$A_i$ and $A'_i$. Each of these would have a different
sensitivity, and thus it would be possible to select the most sensitive
asymmetry for measurement in an experiment. 

For illustration, we have chosen the specific process $e^+e^- \to ZH$, 
for
which we have given explicit expressions for helicity amplitudes, 
density matrices, and the various asymmetries, in the presence of 
anomalous $HZZ$ couplings.

We have also commented on possible determination of one-loop contribution to
triple-Higgs couplings using these asymmetries. 

We hope that the general approach described in detail in this paper will be
found useful to characterize asymmetries and the symmetry properties in
other simple processes of the type $e^+e^- \to Z + X$, with some other 
specific choices of $X$, which may be a single particle, as for example, a
photon, or a pair of particles like Higgs bosons or photons. 

\vskip .3cm
\noindent{\large \bf Acknowledgements} 
SDR thanks the Indian National Science Academy (INSA), New
Delhi, for financial support under the 
Senior Scientist programme.
PS thanks National Science and
Technology Council (NSTC), Taiwan, for financial support.

\vskip .3cm








\end{document}